\documentstyle[12pt]{article}
\begin{document}
\begin{titlepage}
\normalsize
\begin{center}
{\Large \bf Budker Institute of Nuclear Physics} 
\end{center}
\begin{flushright}
BINP 96-49\\

July 1996
\end{flushright}
\vspace{0.5cm}
\begin{center}{\Large \bf  Relativistic corrections }
\end{center}
\begin{center}
{\Large\bf to the positronium decay rate}
\end{center}
\begin{center}
{\Large\bf revisited}
\end{center}

\vspace{0.5cm}

\begin{center}
{\bf I.B. Khriplovich}\footnote{e-mail address: khriplovich@inp.nsk.su}
and {\bf A.I. Milstein}\footnote{e-mail address: milstein@inp.nsk.su}
\end{center}
\begin{center}
Budker Institute of Nuclear Physics, 630090 Novosibirsk, Russia
\end{center}

\vspace{1.0cm}

\begin{abstract}                                    
We rederive here in a simple and transparent way the master formula
for the dominant part of large relativistic corrections to the
positronium decay rate.
\end{abstract}

\end{titlepage}

{\bf 1.} The strong disagreement between the experimental value of the
orthopositronium decay rate \cite{nic}
\begin{equation}
\Gamma^{o-Ps}_{exp}=7.0482(16)\;\mu s^{-1}
\end{equation}
and its theoretical value which includes the order $\alpha$ and
$\alpha^2\log(1/\alpha)$ corrections [2--5]
\begin{equation}\label{th}
\Gamma^{o-Ps}_{th}=m\alpha^6\;\frac{2(\pi^2-9)}{9\pi}\,
\left[1-10.28\frac{\alpha}{\pi}-\frac{1}{3}\alpha^2\log\frac{1}{\alpha}
\right]\,=\,7.03830\;\mu s^{-1}
\end{equation}
is a real challenge to the modern QED. For the disagreement to be
resolved within the QED framework, the correction $\sim
(\alpha/\pi)^2$, which has not been calculated completely up to now,
should enter the theoretical result (\ref{th}) with a numerical factor
250(40), which may look unreasonably large.

Such a hope is not as unreasonable however. One class of large
second-order corrections arises as follows \cite{ky}. The large,
$-10.28$, factor at the $\alpha/\pi$ correction to the decay rate (see
(\ref{th})) means that the factor at the
$\alpha/\pi$ correction to the decay amplitude is roughly 5.
Correspondingly, this correction squared contributes about
$25(\alpha/\pi)^{2}$ to the decay rate. Indeed, numerical calculations
\cite{bu,ad1} give factor $28.86$ at $(\alpha/\pi)^2$ in this
contribution.

One more class of potentially large contributions to the positronium
decay rate is relativistic corrections. A simple argument in their
favour is that the corresponding parameter $(v/c)^2 \sim \alpha^2$
is not suppressed, as distinct from that of usual second-order
radiative corrections, $(\alpha/\pi)^2$, by the small factor
$1/\pi^2 \sim 1/10$.

This problem was addressed in Refs. [9--11].  However, the
discrepancy between the results obtained in \cite{llm}, on one hand,
and in \cite{km,fms}, on another, is huge.  While according to
\cite{llm}, the relativistic correction constitutes (in ``radiative"
units, with $1/\pi$) $24.6(\alpha/\pi)^2$, the results of
\cite{km,fms}, obtained in different techniques, are in a reasonable
agreement between themselves: $46(3)\,(\alpha/\pi)^2$
\cite{km} and $41.9\,(\alpha/\pi)^2$ \cite{fms}. The main source of
this discrepancy can be traced back to the different 
treatment of $(v/c)^2$ arising in the expansion of the
annihilation kernel. While the prescription of \cite{llm} is
effectively (see their formulae (12), (18))
\begin{equation}\label{1/4}
(v/c)^2\,\longrightarrow \,-\,\alpha^2/4,
\end{equation}
our master formula is
\begin{equation}\label{3/4}
(v/c)^2\,\longrightarrow \,-\,3\alpha^2/4.
\end{equation}
Let us mention here that this our recipe refers to the relativistic
correction to the annihilation kernel itself; as to the phase
space correction, we use in it the prescription (\ref{1/4}).

In view of the mentioned discrepancy, we believe that it is
instructive to present an alternative derivation of formula
(\ref{3/4}), more obvious and transparent than our original one.

{\bf 2.} When calculating the decay amplitude, we have to integrate
the annihilation kernel $M(\vec p)$ over the distribution of the
electron and positron three-momenta $\vec{p}$. We address in this
note the relativistic corrections to $M(\vec p)$ only, i.e., we take
as the ground-state wave function $\psi(p)$ the nonrelativistic one.
Then the decay amplitude is
\begin{equation}\label{amp}
\int\,\frac{d\vec p}{(2\pi)^3}\,\psi(\vec{p})\,M(\vec p)\,=\,
\int\,\frac{d\vec p}{(2\pi)^3}\,
\frac{8\sqrt{\pi a^3}}{(p^2 a^2 +1)^2}\,M(\vec p),
\end{equation}
where $a=2/m\alpha$ is the positronium Bohr radius.  To lowest
approximation in $v/c$ the kernel $M(0)$ is independent of those
momenta and we are left with
\begin{equation}\label{amp0}
M(0)\int\,\frac{d\vec p}{(2\pi)^3}\,
\frac{8\sqrt{\pi a^3}}{(p^2 a^2 +1)^2}\,=\,M(0)\,\psi(\vec r=0).
\end{equation}
Thus, in the limit $p\rightarrow 0$ we obtain the common
prescription: the positronium decay rate is proportional to
$|\psi(r=0)|^2$. 

However, already to first order in $(p/m)^2$ the momentum integral
\begin{equation}
\int d\vec{p}\, (p/m)^2 \frac{8\sqrt{\pi a^3}}{(p^2 a^2 +1)^2} 
\end{equation} 
linearly diverges at $p\rightarrow\infty$, which precludes the
straightforward evaluation of these relativistic corrections.

The crucial observation is that the true relativistic expression for
the annihilation kernel does not grow up at $p\rightarrow\infty$, as
distinct from its expansion in $p/m$. So, the initial integral
(\ref{amp}) in fact converges.

When treating relative corrections to the decay amplitude, it is
convenient to single out from it the factor $\psi(r=0)=(\pi
a^3)^{-1/2}$, and a trivial overall dimensional factor from $M(\vec p)$.
So, from now on we investigate, instead of (\ref{amp}), the following
expression:
\begin{equation}\label{amp1}
\int\,\frac{d\vec p}{(2\pi)^3}\,
\frac{8\pi a^3}{(p^2 a^2 +1)^2}\,M(\vec p)
\end{equation}
with dimensionless $M(\vec p)$.

Let us consider first an auxiliary integral
\begin{equation}\label{aux}
\int\,\frac{d\vec p}{(2\pi)^3}\,
\frac{8\pi a^3}{p^4 a^4}\,[M(\vec p)-M(0)],
\end{equation}
which converges both at low and high $p$.
After the angular integration, the dimensionless kernel $M(\vec p)$
depends on the ratio $y^2=(p/m)^2$ only, and the expression
(\ref{aux}) reduces to
\begin{equation}\label{aux1}
\frac{4\alpha}{\pi}\,\int_0^{\infty}\,\frac{dy}{y^2}\,[M(y^2)-M(0)].
\end{equation}
In other words, this auxiliary integral is of first order in
$\alpha/\pi$, and therefore of no interest for our problem. This is a
first-order radiative correction absorbed already by
$-10.28\alpha/\pi$ in (\ref{th}) (for orthopositronium). In fact, we
have neglected in this argument the kernel dependence on the
positronium binding energy (it certainly exists at least in the
noncovariant perturbation theory we were starting from in \cite{km}).
But corrections effectively neglected in this way, are of higher odd
powers in $\alpha$.

So, expression (\ref{aux}) can be used as a regulator, and in the
now rapidly converging integral
\begin{equation}\label{}
\int\,\frac{d\vec p}{(2\pi)^3}\,8\pi a^3
\left[\frac{1}{(p^2 a^2 +1)^2}\,-\,\frac{1}{p^4 a^4}\right]\,
[M(\vec p)-M(0)]
\end{equation}
we can safely expand $M(\vec p)$ up to $(p/m)^2$ included. In this
way we obtain
\begin{equation}\label{}
\int\,\frac{d\vec p}{(2\pi)^3}\,8\pi a^3
\left[\frac{1}{(p^2 a^2 +1)^2}\,-\,\frac{1}{p^4 a^4}\right]\,
\left(\frac{p}{m}\right)^2\,=\,-\,\frac{3}{4}\,\alpha^2. 
\end{equation}
This is an alternative derivation of the master formula (\ref{3/4})
used in our article \cite{km}. In our opinion, this derivation leaves
no doubts in the correctness of this prescription.

\bigskip
One of us (I.Kh.) is grateful to P. Labelle for discussions.

\end{document}